
\magnification=1200
\baselineskip=13pt
\overfullrule=0pt
\tolerance=100000

\def\sqr#1#2{{\vcenter{\vbox{\hrule height.#2pt
        \hbox{\vrule width.#2pt height#1pt \kern#1pt
          \vrule width.#2pt}
        \hrule height.#2pt}}}}
\def\square{\mathchoice\sqr68\sqr68\sqr{4.2}6\sqr{2.10}6}

\def\boxit#1{\vbox{\hrule\hbox{\vrule\kern3pt
   \vbox{\kern3pt#1\kern3pt}\kern3pt\vrule}\hrule}}
\def\lsim{<\kern-2.5ex\lower0.85ex\hbox{$\sim$}\ }
\def\rsim{>\kern-2.5ex\lower0.85ex\hbox{$\sim$}\ }

\rightline{UR-1363\ \ \ \ \ \ \ }
\rightline{ER-40685-813}

\bigskip

\baselineskip=18pt

\centerline{\bf ON THE DERIVATIVE EXPANSION AT FINITE
                TEMPERATURE}

\bigskip
\bigskip

\centerline{Ashok Das*}
\centerline{Instituto de F\'isica}
\centerline{Universidade Federal do Rio de Janeiro}
\centerline{Caixa Postal 68528}
\centerline{21945 Rio de Janeiro,Brasil}
\centerline{and}
\centerline{Marcelo Hott$^\dagger$}
\centerline{Department of Physics}
\centerline{University of Rochester}
\centerline{Rochester, NY 14627}

\bigskip
\bigskip
\bigskip
\bigskip

\centerline{\bf Abstract}

\medskip

In this short note, we indicate the origin of nonanalyticity in the method of
derivative expansion at finite temperature and discuss some of its
 consequences.

\vskip 2 truein

\noindent $^*$Permanent address: Department of Physics
 and Astronomy, University of Rochester,
Rochester, N.Y. 14627.

\smallskip

\noindent $^\dagger$On leave
 of absence from UNESP - Campus de
Guaratinguet\'a, P.O. Box 205, CEP : 12.500, Guaratinguet\'a, S.P., Brazil

\vfill\eject

The derivative expansion [1] has been quite profitably
 used in the study of low
energy properties of various quantum field theories [2,3].
It has also been
applied to the study of various two dimensional models [4,5].
 The idea is quite
simple. The effective action resulting from the
integration of a heavy field can be
expanded in powers of momentum or derivatives.
 In practice, this is implemented
as follows. (See refs.1 and 2 for details.) Consider for simplicity, the theory
described by the Lagrangian density
$$ {\cal L} = {1 \over 2} \partial_\mu \phi (x) \partial^\mu \phi  (x)
- {m^2 \over 2} \phi^2 (x)
- {\lambda \over 2}\ B (x) \phi^2 (x) + {\cal L}_0 \eqno(1)$$
where $\phi$ and $B$ are scalar fields with \lq $m$' representing a heavy mass.
${\cal L}_0 (B)$ is the free Lagrangian density associated with $B$ including
possible linear terms.
 (We choose this theory mainly because it has been studied
in detail in connection with the nonanalytic behavior
 at finite temperature.) One
can, of course, integrate out
 the $\phi$ field in the functional integral and obtain
the effective action for the $B$ field as follows.
$$\eqalign{Z &= \int {\cal D}B\ {\cal D} \phi\ e^{iS[\phi,B]}\cr
                    &= \int {\cal D}B \bigg[ \det (G_F^{-1}(k) - \lambda B(x))
\bigg]^{-1/2}
                           \  e^{iS_0 [B]}\cr
                    &= \int {\cal D}B \bigg[ \det (G_F^{-1}(k)) \bigg]^{-1/2}
 \bigg[ \det (1- \lambda G_F(k) B(x))\bigg]^{-1/2}
\ e^{iS_0 [ B]}\cr
                    &= \int {\cal D}B e^{iS_{\rm eff}[B]}\cr}\eqno(2)$$
where the first nondynamical determinant factor
 has been absorbed into normalization
of the path integral and
$$S_{\rm eff} [B] = S_0 [B] + S^\prime [B]\eqno(3)$$
with
$$ S^\prime [B] = {i \over 2}\  {\rm Tr} \ln (1- \lambda G_F (k)
B(x))\eqno(4)$$
One can expand the logarithm in Eq. (4) in powers to write
$$S^\prime [B] = {i \over 2}\
{\rm Tr} \left[ - \lambda G_F (k) B(x) - {\lambda^2 \over 2} G_F (k)
B(x) G_F (k) B(x) + \dots \right]\eqno(5)$$

The momentum dependent factors can be moved through the coordinate
dependent quantities by use of the usual commutation relations and then
Tr (Trace) can be evaluated by integrating over the momentum and the
coordinate variables. Thus, for example, at zero temperature, the quadratic
part of $S^\prime (B)$ can be written as
$$\eqalign{S^\prime_q [B]
&= -{i \lambda^2 \over 4}\  {\rm Tr}\
 [G_F (k) B(x) G_F (k) B(x)]\cr
           &= -{i \lambda^2 \over 4} \ {\rm Tr}\
 \bigg[ {1 \over k^2 - m^2 + i \epsilon} \ B(x)
               {1 \over k^2 - m^2 + i \epsilon}
B(x) \bigg] \cr}\eqno(6)$$
One can move the momentum factors to the left through the use of the identity
$$\eqalign{B(x)\  &{1 \over k^2 - m^2 + i \epsilon} = {1 \over k^2 - m^2 +i
 \epsilon}\ B(x)\cr
&+ {1 \over (k^2 - m^2 + i \epsilon)^2}\  \big[
                   k^2 , B(x) \big] + {1 \over (k^2 - m^2 + i \epsilon)^3}\
                     \big[ k^2,\big[ k^2,B(x) \big] \big]\cr}\eqno(7)$$
with
$$\left[ k^2, B(x) \right] = (\square\  B(x)) + 2ik_\mu
(\partial^\mu B(x))\eqno(8)$$
The integration over the momentum can now be done leaving us with an effective
action that is expressed in powers of the derivatives.

This is the derivative expansion and by
 construction it implies that the effective action
can always be expanded in powers of  derivatives.
It has also been used in some
calculations of effective action at finite temperature [4,6].
 On the other hand, it is by now
well established that Feynman amplitudes
 do become nonanalytic at finite temperature [7-9].
It is, therefore, interesting to ask how
 the nonanalyticity  manifests
itself  at finite temperature
in the derivative expansion and
 what would be the consequences of such
nonanalyticity.

To address this question,
 let us note that at finite temperature, the
propagator has two terms [10].
$$G_F (k) = G_F^0 (k) + G_F^\beta (k)
           = {1 \over k^2 - m^2 + i \epsilon} - 2i
\pi n_B (k_0) \delta  (k^2 - m^2 )\eqno(9)$$
where the bosonic distribution function has the form
 (One can use a more covariant
description, but we ignore this for the present discussion.)
$$ n_B (k_0) = {1 \over e^{\beta |k_0|} -1}\eqno(10)$$
where $\beta$ is the inverse temperature
 in units of the Boltzmann constant. As a
result of this structure of the finite
 temperature propagator, the quadratic part of the
action in Eq. (6) can be written as
$$S^\prime_q [B] = S^{\prime 0}_q [B] + S_q^{\prime \beta}
[B]\eqno(11)$$
where the temperature dependent part of the quadratic action has the form
$$\eqalign{S^{\prime \beta}_q = &-{i \lambda^2 \over 4}\
 {\rm Tr}\  \bigg[ G_F^\beta (k) B(x) G_F^0 (k) B(x) \cr
        &+ G_F^0 (k) B(x) G_F^\beta (k) B(x) + G_F^\beta (k) B(x)
G_F^\beta (k) B(x) \bigg]\cr}\eqno(12)$$

The crucial observation at this
 point is that any function of momentum can be moved
past a coordinate dependent quantity as

$$B(x)f(k) = (f(k-i \partial) B(x))\eqno(13)$$

The parenthesis on the right hand side merely
 emphasizes that the derivatives act only
on $B(x)$. Each term in Eq. (7) can be checked to
correspond to terms in the Taylor expansion of
Eq. (13). Normally, it should not matter whether we use Eq.
 (7) or Eq. (13) if the quantity
of interest is analytic. However, since we are interested
 in studying the nonanalytic
behavior of the effective action, let us use Eq.
 (13) to move the momentum dependent
factors past the coordinate dependent quantities.
 (Another way of saying this is to
assume that we are summing the series in Eq. (7).)
 In this case, the temperature
dependent part of  the quadratic action will become
$$\eqalign{S^{\prime \beta}_q [B(x)]  &= -{i \lambda^2 \over 4}
 \ {\rm Tr}\ \bigg[ G_F^\beta (k) (G_F^0 (k-i \partial)
           B(x)) B(x)\cr
&\qquad + G_F^0 (k) (G_F^\beta (k- i \partial) B(x)) B(x) \cr
&\qquad        + G_F^\beta (k) (G_F^\beta (k-i \partial)B(x)) B(x) \bigg]\cr
    &= - {\lambda^2 \over 4} \int {d^4 k \over (2 \pi)^3} \int d^4 x
       \bigg[  n_B (k_0) \delta (k^2 -m^2) {1 \over (k-i \partial)^2 - m^2 +
     i\epsilon} B(x)\cr
&\qquad  + n_B (k_0 - i\partial_0)  {1 \over k^2 -m^2 + i \epsilon}
\ \delta ((k-i \partial) ^2 -m^2) B(x)\cr
         &\qquad - 2i \pi n_B (k_0) n(
 k_0 -i \partial_0) \delta (k^2 -m^2) \delta ((k- i \partial)^2 -m^2) B(x)
             \bigg] B(x)\cr}\eqno(14)$$
We emphasize again that the derivatives are supposed
 to act only on the first factor
of $B(x)$. We also note that the momentum integral
 in Eq. (14) is nothing other than
the one studied in detail [8,9] in connection
 with the nonanalyticity associated with
the two point function at finite temperature
 (with the identification $p = - i \partial$). In fact,
we can evaluate this using the modified Feynman
 parameterization [9] and show that
$${\rm Re}\  S^{\prime \beta}_q [B] =
 - {\lambda^2 \over 32 \pi^2}  \int d^4x \left\{ \int_0^\infty {kdk \over
     \omega} n_B (\omega)
{1 \over (-\vec \nabla^2)^{1/2}} \ {\rm Re}\ (\ln R) B(x) \right\}
                                     B(x)\eqno(15)$$
where
$$\omega = (k^2 + m^2)^{1/2}\eqno(16)$$
and
$$R = {(\partial_0^2
- \vec \nabla^2  + 2 i \omega \partial_0 + 2ik(-\vec \nabla^2)^{1/2})(
\partial_0^2
          - \vec \nabla^2
-  2 i \omega \partial_0 + 2ik
(-\vec \nabla^2)^{1/2}) \over
 (\partial^2_0 - \vec \nabla^2
+ 2 i \omega \partial_0 - 2ik (-\vec
\nabla^2)^{1/2})(
 \partial_0^2 - \vec \nabla^2 -
 2i \omega \partial_0 -
           2 i k (-\vec \nabla^2)^{1/2})}\eqno(17)$$

The temperature dependent action in Eq. (15)
 is manifestly nonanalytic and does not have
an expansion in powers of $\partial_\mu$.
This is, of course, the nonanalyticity that is
most widely studied. But in principle, the
 cubic, quartic and other terms in the effective
action may develop similar nonanalytic structure.
 We note here that there are examples [11]
where the two point function is analytic, but in
 general nonanalyticity is present at finite
temperature.

This, therefore, shows that the derivative expansion
 really breaks down at finite
temperature. If one uses Eq. (7) to move momentum dependent quantities to the
left of the coordinate dependent quantities,
 then, of course, one has a well defined
derivative expansion. But this corresponds to
Taylor expanding the integrand before
evaluating the integral which coincides with
 Taylor expanding the final action around
the origin obtained in the limit
 $\partial_0 = 0$ and $(\vec \nabla^2)^{1/2} \rightarrow 0$. There
is, of course, no a priori reason why this should
 be the proper limit. In other words,
Taylor expansion before evaluating the integral
is meaningful for an analytic action.
However, when the action is nonanaytic,
as is the case at finite temperature, such
an expansion becomes questionable. In the same spirit,
 we note that since a dervative
expansion of the effective action is not rigorously
 possible, the definition of an effective
potential [12], in such a case, is not unique.
 This, of course, has far reaching
consequences in connection with studies in symmetry breaking
 and restoration [13] and
needs further study.

This work was supported in part by the U.S. Department of Energy Grant No.
DE-FG-02-91ER40685.  One of us (A.D.) would like to thank CNPq, Brazil, for
financial support and the members of the Instituto de F\'isica
 (Universidade Federal do Rio de Janeiro)
for hospitality during
the course of this work. M.H. would like to thank the
 Funda\c c\~ao de Amparo a Pesquisa
do Estado de S\~ao Paulo for financial support.

\vfill\eject

\noindent {\bf References}

\medskip

\item{1.} C.M. Fraser, Z. Phys. {\bf C28}, 101 (1985); V.A. Novikov, M.A.
 Shifman, A.I. Vainshtein and V.I. Zakharov, Sov. J. Nucl. Phys. {\bf 39},
77 (1984).

\item{2.} I.J.R. Aitchison and C.M. Fraser, Phys. Rev. {\bf D31}, 2605 (1985).

\item{3.} I.J.R. Aitchison and C.M. Fraser, Phys. Lett. {\bf 146B}, 63 (1984);
I.J.R. Aitchison and C.M. Fraser, Phys. Rev. {\bf D32}, 2190 (1985); I.J.R.
 Aitchison, C.M. Fraser and P.J. Miron, Oxford Report No. 49/85.

\item{4.} A. Das and A. Karev, Phys. Rev. {\bf D36}, 623 (1987).

\item{5.} A. Das and A. Karev, Phys. Rev. {\bf D36}, 2591 (1987).

\item{6.} Ch. G. van Weert in \lq\lq Proceeding of 2nd Workshop on Thermal
 Field Theories and  Their Applications", ed. by H. Ezawa, T. Arimitsu,
Y. Hashimoto, (North Holland, Amsterdam), p.17 (1990).

\item{7.} O.K. Kalashnikov and V.V. Klimov, Sov. J. Nucl. Phys. {\bf 31},
 699 (1980); H.A. Weldon, Phys. Rev. {\bf D26}, 1394 (1982).

\item{8.} P.S. Gribosky and B.R. Holstein, Z. Phys. {\bf C47}, 205 (1990);
P.F. Bedaque and A. Das, Phys. Rev. {\bf D45}, 2906 (1992).

\item{9.} H.A. Weldon, Phys. Rev. {\bf D47}, 594 (1993); P.F. Bedaque and
A. Das, Phys. Rev. {\bf D47}, 601 (1993).

\item{10.} One can use any real time description for this. See for example,
J. Schwinger, J. Math. Phys. {\bf 2}, 407 (1961); J. Schwinger, Lect. Notes of
Brandeis Summer Institute in Theoretical Physics (1960); L.V. Keldysh, Sov.
Phys. JETP {\bf 20}, 1018 (1968); P.M. Bakshi and K.T. Mahanthappa, J. Math.
Phys. {\bf 4}, 1 (1963); H. Umezawa, H. Matsumoto and M. Tachiki,
\lq\lq Thermo Field Dynamics and Condensed States", North Holland (1982);
L. Dolan and R. Jackiw, Phys. Rev. {\bf D9}, 3320 (1974).

\item{11.} P. Arnold, S. Vokos, P.F. Bedaque and A. Das, Phys. Rev. {\bf D47},
 498 (1993).

\item{12.} See any standard text on Quantum Field Theory for a definition
of the effective potential. For example, A. Das, \lq\lq Field Theory:
 A Path Integral
Approach", p.249, World Scientific (1993).

\item{13.} See for example, the last article in ref. 10.

\end